\documentstyle[12pt]{article}
\def\beq{\begin{equation}}   \def\eeq{\end{equation}}
\def\bea{\begin{eqnarray}}   \def\eea{\end{eqnarray}}

\begin{document}

\begin{flushright}
UND-HEP-00-BIG\hspace*{.2em}04\\
June 2000\\
\end{flushright}
\vspace{.3cm}
\begin{center} \Large 
{\bf Dear Santa: Heavy Flavour Physics at 
Neutrino Factories -- Desires from Theory}
\footnote{Talk given at NuFact'00, May 22 - 26, 2000,
Monterey, CA}
\\
\end{center}
\vspace*{.3cm}
\begin{center} {\Large 
I. I. Bigi }\\ 
\vspace{.4cm}
{\normalsize 
{\it Physics Dept.,
Univ. of Notre Dame du
Lac, Notre Dame, IN 46556, U.S.A.} }
\\
\vspace{.3cm}
{\it e-mail address: bigi@undhep.hep.nd.edu } 
\vspace*{1.4cm}

{\Large{\bf Abstract}}
\end{center}

Even in 2010 the CKM parameters $V_{cs}$, $V_{cd}$ and 
$V_{ub}$ will be known with less than desirable 
accuracy; the discovery potential for New Physics in 
charm decays -- in particular their CP asymmetries -- 
will be far from
exhausted; important tests of our theoretical tools will 
not have been performed. I sketch the impact a $\nu$fact could 
have in these areas. 

\vspace*{.2cm}
\vfill
\noindent
\vskip 5mm


During this talk I will attempt to sketch which important information 
on heavy flavour physics will still be missing in 2010. My personal 
crystal ball tells me that the knowledge base available at that time 
can be enlargened in three aspects and that neutrino factories 
($\nu$fact) might be up to the task eventually: 
\begin{itemize}
\item 
Some basic SM quantities will be known with less than desirable 
accuracy. 
\item 
The discovery potential in charm decays will be far from 
exhausted. 
\end{itemize} 
Realizing even a single item in these categories through a 
new initiative would provide a strong motivation for the latter. 
However to make a conclusive case that some fundamental parameter 
had indeed been determined more reliably or that the intervention of 
New Physics had been revealed, one has to make sure that our 
{\em theoretical} tools are up to the task: 
\begin{itemize}
\item 
The need might still exist to test our theoretical 
technologies in charm decays. 

\end{itemize}
Obviously one or two measurements are not sufficient to do the 
job here -- a broad and detailed program is called for. 

\section{Basic Quantities} 

PDG2000 quotes the following errors on $V_{cb}$ and $V_{ub}$: 
\beq 
|\Delta V_{cb}| \; \hat = \; 8 \% \; \; \; , \; \; \; 
|\Delta V_{ub}| \;  \hat = \; 40 \%
\eeq
For $V_{cs}$ and $V_{cd}$ two set of errors are listed: 
\beq 
|\Delta V_{cs}| \; \hat = \; 17 \% \; [2\%]\; \; , \; \; \; 
|\Delta V_{cd}| \;  \hat = \; 20 \% \; [3\%]
\eeq
where the first numbers refer to {\em direct} extractions 
and the second ones in square brackets reflect what happens 
upon imposition of three-family unitarity. $|V_{cs}|$ and 
$|V_{cd}|$ have been studied in semileptonic $D$ decays as well 
as in neutrino production of charm with the former having 
more weight in $|V_{cs}|$ and the latter in $|V_{cd}|$. 

My expectation is that these uncertainties will be reduced 
significantly over the next decade, albeit not by an 
order of magnitude: 
\bea 
|\Delta V_{cs}|_{pre-\nu fact} &\sim& 10 \% \; [2\%], \; 
|\Delta V_{cd}|_{pre-\nu fact} \sim 10 \% \; [2\%],  \\ 
|\Delta V_{cb}|_{pre-\nu fact} &\sim& 4 \%, \; 
|\Delta V_{ub}|_{pre-\nu fact} \sim 10 - 15 \%, \;
\eea 
with the quoted uncertainty of 10 - 15 \% in 
$|V_{ub}|$ not being guaranteed \cite{GIBBONS}. 

There are strong reasons why we want to reduce 
these uncertainties further still: 
\begin{itemize}
\item 
The CKM parameters are fundamental quantities related to a  
central mystery of the SM, namely the generation of fermion 
masses.  
Many intriguing suggestions have been made to explain these 
parameters in terms of so-called `textures' assumed to hold among 
Yukawa couplings at GUT scales. However even if those texture 
patterns look completely different at GUT scales, those differences tend 
to get substantially diminished when running the couplings down 
to electroweak scales where they can be probed. 
\item 
By 2010 various CP asymmetries in $B$ decays should be 
measured with errors of about very few percent. 
One pre-requisite for a matching accuracy in the 
KM prediction is to know CKM parameters to very few percent as well. 

\end{itemize} 
A high rate $\nu$fact might be harnessed 
\cite{BNL} to provide a competitive or 
even superior determinations of some CKM parameters -- as is 
at present the case with $|V_{cs}|$ and $|V_{cd}|$-- through 
measuring the heavy flavour {\em production} 
cross section off the appropriate quarks. 
Coupling the high statistics with a high quality 
vertex detector should enable one to measure charm production 
with an accuracy below 1 \%. The problem is 
to which degree one can predict such a cross section 
as a function of $V_{cs}$ and $V_{cd}$. Since one is comparing 
$d[s] \to c$ with $d[s] \to u$, uncertainties in the 
parton distribution functions will drop out. The central 
problem is to which degree of accuracy one can deal 
quantitatively with the suppression of charm (and 
ultimately beauty) quark production. A 20 \% accuracy 
in the cross section translating into a 10\% accuracy in 
$V_{cd[s]}$ should be achievable, yet one wants to aim higher. 

There are two major theoretical stumbling blocks: 
(i) The models one has been using to describe the on-set of 
charm production are of a purely phenomenological nature. 
The quark mass parameter they contain is not related to 
the charm mass properly defined in QCD. With the threshold 
region shaped by non-perturbative dynamics, one needs a 
nonperturbative definition of the quark mass. 
While such a definition 
has been developed for heavy flavour decays 
\cite{rev}, this has not 
happened yet for threshold production. 
(ii) Uncertainties in the charm fragmentation function 
constitute a serious limiting factor. 

There are two avenues that in combination might lead us 
towards better theoretical control over charm production: on one hand 
one can undertake to carry over technologies developed to deal with 
non-perturbative dynamics in heavy flavour decays to describe heavy
flavour  production; on the other hand 
one can analyze to which degree 
an experiment at such a $\nu$fact could measure 
fragmentation effects with sufficient accuracy to reduce the
aforementioned theoretical uncertainties. I would conclude 
that extracting $|V_{cs}|$ and $|V_{cd}|$ within 10 \% 
should be achievable with the hope that based on further 
theoretical and experimental work those uncertainties can be 
significantly reduced. Furthermore the aforementioned 
complications should basically drop out from the ratio 
$|V_{cd}/V_{cs}|$.

Less fundamental, yet very instrumental is the extraction of the 
decay constants $f_{D}$ and $f_{D_s}$ from 
$D \to \mu \bar \nu$ and $D_s \to \mu \bar \nu$, respectively. 
For comparing the measured value with the one predicted by, say, 
lattice simulations of QCD would  
provide an important
calibration  of that methodology; if it passes, one would be much 
more confident in its extrapolation to $f_B$, the analoguous
quantity  for $B$ mesons. Having a reliable value for the 
latter, one could then infer $|V_{td}|$ from 
$\Delta M_B$. 
Identifying such 1-prong channels by their kink constitutes
a  formidable challenge. An excellent vertex resolution in a 
clean environment might provide us with the answer to this 
challenge. 

\section{Testing our 
Theoretical Tools in Charm Decays}
The expectation that detectors at a 
$\nu$fact can function like an {\em electronic bubble chamber} in a 
clean environment gives rise to the hope that measurements 
can be performed here that otherwise require a dedicated 
charm factory. 

Very little  
is known about the {\em absolute} 
values of branching ratios for the various charm baryons;  
the situation for $D_s$ is only somewhat better, and the status 
for $D^+$ and $D^0$ is nothing to brag about, either. 
With these absolute 
values representing important engineering input into studies of 
$B$ decays, they have recently emerged as one of the 
limiting factors. A high quality detector at a 
$\nu$fact might allow us to infer absolute branching ratios. In the 
absence of a charm factory the only 
other method known relies on $B$ decays as source for tagged 
charm hadrons \cite{NELSON}. 

The next step would be measuring 
absolute branching ratios of {\em inclusive} 
semileptonic $D_s$, $\Lambda _c$, $\Xi _c^{+,0}$ and 
$\Omega _c$ decays. The $1/m_c$ expansion makes some 
highly non-trivial predictions here based on the 
occurance of sizeable constructive interference 
effects in the semileptonic widths \cite{VOLOSHIN}. 
Whether $1/m_c$ expansions hold or not is an important 
issue in its own right and a crucial element in an analysis 
of $D^0-\bar D^0$ oscillations. 

Finally measuring charm transition rates into 
multi-neutral final states provides us with lessons on 
quark-hadron duality at the charm 
scale. It will also help to interprete properly CP asymmetries 
once they are observed in $D$ decays. 

\section{Identifying New Physics in Charm Decays}
$D^0-\bar D^0$ oscillations are driven by 
the normalized mass and width 
differences:
\beq 
x_D \equiv \frac{\Delta M_D}{\Gamma _D} \; \; , \; \; 
y_D \equiv \frac{\Delta \Gamma}{2\Gamma _D} 
\eeq
I share the usual expectation that while $x_D$ is naturally  
sensitive to New Physics, $y_D$ is not (except for some 
contrived scenarios). 

The usual folklore is based on two statements: (i) The 
contributions from the quark box diagrams  
are highly suppressed and insignificant. (ii) Long 
distance dynamics yield the leading contributions 
with $x_D$, $y_D$ $\sim 10 ^{-4} - 10^{-3}$. New Physics 
could naturally enhance $x_D$ to the few percent level. 
This might be described as the `King Kong' scenario: 
one is unlikely to ever encounter King 
Kong; yet once it happens there can be no doubt that one has 
come across someting out of the ordinary.

A recent careful SM analysis leads to the following 
conclusions \cite{BUDD}: The operator product expansion 
provides a
coherent and self-consistent description.  The $SU(3)$ suppression of the
box contributions  described by $m_s^4/m_c^4$ is untypically 
strong. Other contributions (given by 
quark condensates) with GIM factors 
$m_s^2/m_c^2$ or even $m_s/m_c$ are numerically leading. 
There is no need 
to postulate additional long distance contributions. 
The numerical estimates, however, change little: 
\beq 
x_D\, , \; y_D \; \sim {\cal O}(10^{-3})  
\eeq
Studying oscillations requires 
a flavour tag both in the initial and the final state. 
So far mainly $D^{*,+} \to D^0 \pi ^+$ vs. 
$D^{*,-} \to \bar D^0 \pi ^-$ have been used for 
initial state tagging. A $\nu$fact would naturally 
use the muon of the CC interaction as the initial 
flavour tag. The final state flavour can be 
identified by its strangeness or its lepton number:
$D^0 \to K^- \pi ^+$ or $D^0 \to l^+ X$. Whereas 
there is no SM background to the latter, there is one 
to the former, namely the doubly Cabibbo 
suppressed mode. Studying the time evolution of the 
decay rate allows one to separate out that component: 
$$ 
{\Gamma}( D^0(t) \to K^+ \pi ^-) \propto 
e^{-\Gamma _{D^0} t} {\rm tg}^4\theta _C|\hat{ \rho }_{K\pi }|^2 
$$ 
$$ 
\times \left[ 1 - \frac{1}{2}\Delta \Gamma _D t + 
\frac{(\Delta m_Dt)^2}{4{\rm tg}^4\theta _C
|\hat{\rho}_{K\pi }|^2} 
+ \frac{\Delta \Gamma _Dt}
{2{\rm tg}^2\theta _C|\hat{\rho }_{K\pi }|}
{\rm Re}\left( 
\frac{p}{q}\frac{\hat{\rho }_{K\pi }}
{|\hat{\rho }_{K\pi }|}   
\right) \right. 
$$ 
\beq
\left. - \frac{\Delta m_Dt}{{\rm tg}^2\theta _C
|\hat{\rho }_{K\pi }|} 
{\rm Im}\left( 
\frac{p}{q}\frac{\hat{\rho }_{K\pi }}{|\hat{\rho }_{K\pi }|}   
\right)  
\right] \; , \; \; 
{\rm tg}^2\theta _C \cdot \hat{\rho }_{K\pi } \equiv 
\frac{T(D^0 \to K^+ \pi ^-)}{T(D^0 \to K^- \pi ^+)}
\label{DKPI}
\end{equation}
Another observable is the integrated rate into `wrong-sign' 
leptons: 
\beq 
r_D \equiv \frac{{\rm rate} (D^0 \to l^- X)}
{{\rm rate} (D^0 \to l^+ X)} \simeq 
\frac{1}{2}\left( x_D^2 + y_D^2\right)  
\eeq  
Furthermore one can compare the lifetimes determined from 
different channels, like $D^0 \to K^+ K^-$ vs. 
$D^0 \to K^- \pi ^+$. The experimental landscape can be 
sketched by the following numbers \cite{NELSON}: 
\beq 
|x_D|, \; |y_D| \; \leq 0.028\; , \; 
- 0.058 \leq y_D^{\prime} \leq 0.01 \; , \; \; 
95 \% \; {\rm C.L.\; \; CLEO} 
\eeq 
where 
$y_D^{\prime} \equiv y_D {\rm cos} \delta _{K\pi} - 
x_D {\rm sin} \delta _{K\pi}$  
with $\delta _{K\pi}$ denoting the strong phase shift 
between $D^0 \to K^+ \pi ^-$ and $D^0 \to K^- \pi ^+$;  
\beq 
y_D = 0.0342 \pm 0.0139 \pm 0.0074 \; \; \; 
FOCUS
\eeq 
{\em If} FOCUS has seen a genuine signal, then we find 
ourselves in a conundrum: the value for $y_D$ is an 
order of magnitude larger than expected -- yet it can hardly 
be attributed to New Physics! While it suggests 
that $\Delta M_D$ is `just around the corner', it would force us 
to abandon the `King Kong' scenario: it makes any claim of 
New Physics based merely on the observation of oscillations 
of very dubious validity. 

It is expected that the $B$ factories can probe 
$x_D$ and $y_D$ down to just below 1\% \cite{NELSON}. 
That means that there is a good chance that the 
question of $D^0 - \bar D^0$ oscillations has not been fully 
answered on the experimental level in 2010. It 
remains to be seen whether a $\nu$fact could go down even further 
by combining different decay channels in the analysis. This 
involves also issues of systematics. One should  
probe decays into wrong-sign leptons for the presence of 
a prompt {\em non-SM} component through 
analysing the time evolution in analogy to 
Eq.(\ref{DKPI}). 


Doubly Cabibbo suppressed (DCS) decays of neutral $D$ mesons are a 
promising area to search for CP violation. While one pays a 
heavy price in statistics, the asymmetry can get much larger 
since it involves the interference between the DCS amplitude 
and the oscillation amplitude. The time evolution for 
$\bar D^0(t) \to K^-\pi^+$ is obtained from the one for 
$D^0(t) \to K^+\pi^-$, see Eq.(\ref{DKPI}), by substituting 
the analoguous amplitude ratio 
tg$^2\theta _C \bar \rho _{K\pi}$ for 
tg$^2\theta _C  \rho _{K\pi}$ and flipping the sign of the last term. 
This CP asymmetry is controlled by $x_D/{\rm tg}^2\theta _C$. 
{\em If} $x_D \simeq \frac{1}{2}\%$ -- thus possibly beyond 
the reach of the beauty factories -- one had 
$x_D/{\rm tg}^2\theta _C \simeq 0.1$; i.e., the CP asymmetry 
could conceivably be as large as up to 10\% {\em without} 
oscillations having been found through CP insensitive observables 
$\propto x_D^2$! It is important to analyze how small an asymmetry 
could be found at a $\nu$fact. With a sample size of 
$10^8$ charm hadrons one should reach the 1\% level here 
statistically; yet it is more
than a question  of statistics. A similar exercise should be done
for 
$D^0(t) \to K^+K^-$. Since the final state is a CP eigenstate, 
its time evolution is less complex than for the previous case 
\cite{BNL}. 
In any case, we can be quite confident that any such signal 
that could be found would reveal the intervention of New 
Physics -- unlike the situation with CP 
{\em insensitive} oscillation 
observables. 

The KM ansatz allows for {\em direct} CP 
asymmetries to emerge  
in some Cabibbo suppressed channels plausibly reaching the 
${\cal O}(10^{-3})$ level. If such an effect were observed 
on the 0.1\% or even 1\% level, one would like to decide whether 
it was still compatible with the KM ansatz or required the 
intervention of New Physics. An important element in such 
an analysis would be to reach the required 
experimental sensitivity in several channels, in particular 
those that contain neutrals to determine the size and 
phases of the contributing isospin amplitudes.  

\section{Summary}
The urgency for obtaining answers to some open questions in 
heavy flavour decays might actually have increased in 2010: 
(i) 
With the experimental errors for CP asymmetries in $B$ 
decays having been decreased to about 2\% a matching accuracy in 
the predicted values will be desirable. 
(ii)  
If $D^0 - \bar D^0$ oscillations and CP asymmetries in charm 
decays had been found, one would 
like to know whether they require New Physics.

It appears possible that a $\nu$fact with its novel 
and even superior systematics could make substantial
contributions to these areas: 
they might allow more sensitive measurements of either 
the primary effect or of secondary effects that would help us 
in a proper interpretation of the observations. 

\vskip 3mm  
{\bf Acknowledgements} 
  
The organizers deserve heartfelt thanks to finding such a lovely 
site. 
This work has been supported by the NSF under the grant 
PHY 96-05080. 


\end{document}